# Efficient narrow-band light emission from a single carbon nanotube p-n diode

Thomas Mueller[1*†], Megumi Kinoshita[1,2†], Mathias Steiner[1], Vasili Perebeinos[1], Ageeth A. Bol[1], Damon B. Farmer[1], and Phaedon Avouris[1*]

[†]*These authors contributed equally to this work.*
[*]*Email: thomas.mueller@tuwien.ac.at, avouris@us.ibm.com*

[1]IBM Thomas J. Watson Research Center, Yorktown Heights, NY 10598, USA
[2]Department of Physics and Astronomy, Stony Brook University, Stony Brook, NY 11794, USA

**Electrically-driven light emission from carbon nanotubes**[1–8] **could be exploited in nano-scale lasers**[9] **and single-photon sources**[10]**, and has therefore been the focus of much research. However, to date, high electric fields and currents have been either required for electroluminescence**[4–8]**, or have been an undesired side effect**[2,3]**, leading to high power requirements and low efficiencies. In addition, electroluminescent linewidths have been broad enough to obscure the contributions of individual optical transitions. Here, we report electrically-induced light emission from individual carbon nanotube p-n diodes. A new level of control over electrical carrier injection is achieved, reducing power dissipation by a factor of up to 1000, and resulting in zero threshold current, negligible self-heating, and high carrier-to-photon conversion efficiencies. Moreover, the electroluminescent spectra are** **significantly narrower (~35 meV) than in previous studies**[1–8]**, allowing the identification of emission from free and localized excitons.**

P-n junction diodes are the basic building blocks of almost all of today's optoelectronic devices, such as photo-detectors, light emitting diodes (LEDs), and lasers. The demonstration of light emission from carbon nanotube p-n diodes is thus a fundamental step towards a possible technological use of nanotubes as nanometer-scale light sources. Here we provide the first such demonstration. In our nanotube diodes (Fig. 1a), the p- and n-type regions are formed using the electrostatic doping technique introduced by Lee and co-workers[11]. For details, we refer to the Methods section.

In Fig. 1b we present the IV characteristics of a single-walled carbon nanotube diode under two different biasing conditions. The dashed green curve shows the drain-source current $I_{DS}$ versus drain-source voltage $V_{DS}$ when both gate biases are negative: $V_{GS1} = V_{GS2} = -8$ V. The tube then behaves as p-type resistor and a symmetric, almost ohmic conduction behavior is observed. The deviation from a completely linear IV characteristic (dotted curve) at low $V_{DS}$ is attributed to a voltage drop at the Schottky contacts between the metal electrodes and the nanotube. By applying gate biases of opposite polarity a p-i-n diode is realized. The solid red line in Fig. 1b shows the IV characteristic recorded with $V_{GS1} = -8$ V and $V_{GS2} = +8$ V. The device now clearly shows rectifying behavior. The corresponding bandstructure is shown in Fig. 1c.

The left image in Fig. 2a depicts the infrared emission from a device when the gate electrodes are biased at $V_{GS1} = -8$ V, $V_{GS2} = +8$ V and a constant current of $I_{DS} = 240$ nA is driven through the tube. In general, electrically excited light emission from semiconducting carbon nanotubes can be produced under (i) ambipolar[2,3] or (ii) unipolar[4,5] operation. In the first case, both electrons and holes are injected simultaneously into the tube and their radiative recombination generates light. In the second case, a single type of carriers, i.e. either electrons or holes, accumulate kinetic energy in a high-field region within the device to generate excitons by means of impact excitation. The fact that no light is emitted when our devices are operated under unipolar conditions ($V_{GS1} = V_{GS2} = -8$ V; hole current) – right image in Fig. 2a – shows that they are ambipolar light-emitters. This is the behavior we would generally expect a LED to exhibit. The signal is still detectable at $I_{DS}$ as low as ~10 nA. This is in contrast to all previous electroluminescence (EL) studies[1–8], where typically two orders of magnitude higher current levels are required to obtain light emission of comparable intensity. Moreover, the voltage drop across the intrinsic region is in the order of the bandgap (~1 V; see Fig. 1b), and therefore also 5–10 times smaller than in other devices[1–8], overall resulting in an up to 1000 times smaller power dissipation. Under typical operation conditions, we estimate a power density of only ~0.1 W/m in the tube, compared to the 10–100 W/m in other devices. It is hence clear that the nanotube diodes are operated in an entirely different regime than all other electrically-driven carbon nanotube light-emitters to date. In fact, the power density is comparable to what is typically used in photoluminescence (PL) experiments[12], and thermal heating, which strongly influences the EL of metallic as well as semiconducting nanotubes[7], does not play a role.

After calibrating our detection system against the infrared emission from a known black-body emitter and taking into account its collection efficiency (see Supplementary information), we estimate an EL efficiency of ~0.5–1 × $10^{-4}$ photons per injected electron-hole pair. Given a radiative carrier lifetime of 10–100 ns in nanotubes[13–15], we hence obtain a non-radiative lifetime $\tau_L$ in the order of a few picoseconds. This value is smaller than what is typically observed in PL measurements[16] (~20–200 ps). It appears reasonable, though, because of the interaction with the environment and the higher carrier concentrations, which both cause an increase of the non-radiative decay rate[17]. We also acquired the EL spectra of our devices (for details see Methods section). Fig. 2b shows the spectrally dispersed emission of a nanotube diode at $I_{DS}$ = 200 nA. It is composed of a single, narrow peak centered at ~0.635 eV, with a spectral width of ~50 meV (full width at half maximum - FWHM). Based on a correlation between the EL results with PL and Raman data[18], we assign the EL peak to emission from the lowest-energy bright exciton state $E_{11}$ in the nanotube (see Supplementary information).

In Fig. 3a we present the results obtained from another device. Besides the dominant emission at ~0.755 eV (labeled *X*), a weaker luminescence band is observed at ~65 meV lower energy (*LX*). We can rule out the possibility that *X* and *LX* are originating from two separate tubes in a multi-walled carbon nanotube. The small energy spacing translates into a diameter difference which is much less than twice the graphite lattice-plane distance. In order to further confirm that both emission peaks do not stem from a small bundle of nanotubes, we characterized the tube by resonance Raman spectroscopy and atomic force microscopy (AFM). In the Raman measurements, we tune the excitation laser energy between 2.0 and 2.5 eV, i.e. across the $E_{33}$-range that corresponds to the diameter-range of our sample (see Supplementary information). Only one single radial-breathing-mode (RBM) centered at $\Omega_{RBM}$ ~ 200 cm$^{-1}$ is observed, from which the nanotube diameter is determined to be[19] $d_t = 248/\Omega_{RBM}$ ~ 1.24 nm. From the AFM cross-section we extract a similar diameter. Those measurements, and the fact that double-peak spectra similar to the one in Fig. 3 have also been observed in other devices, support our claim that both emission features originate from a single-walled tube. Upon decreasing the gate bias voltages from -/+ 9 V to -/+ 7 V, the doping in the p- and n-regions decreases and so does the infrared emission intensity (for same $V_{DS}$). However, as shown in Fig. 3b, we also observe an even further reduction of the spectral width. We now extract a linewidth of only ~35 meV (FWHM). This value is ~2–8 times smaller than what has been reported in all previous EL measurements to date[1–8], and approaches that typically observed in room-temperature PL[12] (~25 meV).

The two peaks in Fig. 3 cannot be identified as the $E_{11}$ exciton transition and the $E_{11}$ continuum. The exciton binding energy that we estimate for a 1.24-nm-diameter tube embedded in SiO$_2$/Al$_2$O$_3$ ($\varepsilon_{eff}$ ~ 5.7) is[20] $E_b$ ~ 0.12 eV, i.e. almost twice as large as the observed splitting. More importantly, the continuum transition carries only a small fraction of the spectral weight[20] (see also Fig. 4). We can also exclude phonon-assisted emission, because of the different current-dependencies of the two peaks. Low-energy satellite peaks have repeatedly been observed in PL measurements and have been attributed to localized exciton states[10,16,21,22]. We thus assign the peak *X* to "free" exciton emission and *LX* to emission from weakly localized excitons. It is not possible to determine from our optical measurements the physical mechanism of the exciton localization. It might be due to environmental fluctuations, leading to the formation of quantum-dot-like states, or brightening of intrinsic dark states at structural defect sites[14]. We note that in one of our devices, the low-energy emission feature *LX* was initially not present, but developed after stressing the tube by passing a high current through the device. This observation supports the assignment of *LX* to emission from a defect site.

Fig. 3c depicts the current dependence of the $X$ and $LX$ emission intensities as extracted from Fig. 3a. The free exciton emission $X$ shows a linear increase with current. This is in contrast to previous studies of EL from nanotubes[1–8], that exhibit current thresholds of >1 µA for light emission. The nanotube diodes thus constitute threshold-less nano-scale light-emitters. The localized exciton $LX$ rises linearly at low currents but saturates as the current exceeds ~100 nA. Saturation of exciton emission in nanotubes is a characteristic signature of Auger-mediated exciton-exciton annihilation[23–25], which is known to be strongly enhanced in tightly confined 1D systems[26]. It sets in when more than one exciton is present in the tube, i.e. when the electron-hole pair injection rate $I_{DS}/2q$ ($q$ is the electron charge) exceeds the inverse carrier lifetime $\tau_L^{-1}$. Therefore, $\tau_L = 2q/100$ nA ~ 3 ps, which is in agreement with the $\tau_L$ estimated from the EL efficiency above. The sudden saturation further suggests that $\tau_A^{LX} \ll \tau_L$, with $\tau_A^{LX}$ being the $LX$–$LX$ annihilation lifetime. From the absence of any noticeable $X$ saturation, on the other hand, we expect the $X$–$X$ annihilation lifetime $\tau_A^X$ to be much longer than $\tau_L$. In fact, following Ref. 26, we estimate ($E_b = 0.12$ eV, $E_g = 0.755$ eV + $E_b$) $\tau_A^X$ ~ 12 ps. The fact that the emission from defect sites is of comparable strength as the emission from the rest of the tube further points towards a strong exciton nucleation (i.e. locally increased exciton density) at the low-energy defect sites. Fig. 3d depicts the results from another device. A similar behavior is observed, but the onset of saturation now occurs at higher current (~250 nA), suggesting a higher concentration of defects in this tube. This is also consistent with a ~2 times stronger $LX$ emission as compared to $X$ emission at low currents. It might also be interesting to investigate the photon statistics of the $LX$ emission, since it is believed that exciton localization plays an important role in the generation of quantum light from nanotubes[10].

In Fig. 4 we compare the ambipolar emission from a diode with the unipolar emission from a back-gated field-effect transistor (FET) made out of the same, long tube. The FET emission amplitude is ~4 times smaller than the emission from the diode, although $I_{DS}$ is 12.5 times higher. It is also spectrally broader (~180 meV) and exhibits a slightly asymmetric lineshape. The FET device was operated in the reverse bias regime, with $V_{GS} < 0 < V_{DS}$ and $|V_{GS}| > |V_{DS}|$. In this regime, holes are the majority carriers and generate electron-hole pairs by impact excitation. Most electron-hole pairs are generated at the peak field $F_{max}$ near the drain electrode (see inset) and we estimate a lower limit of $F_{max} > V_{DS}/t_{ox} \sim 25$ V/μm, where we use the gate oxide thickness $t_{ox}$ as the screening length[27]. When estimating the contribution of different broadening mechanisms to the emission linewidth we find that, under those biasing conditions, the dominant contribution is due to mixing of exciton and continuum states in the high electric field. The inset of Fig. 4 shows a simulation[28] of the field dependence of the optical absorption of a 1.4 nm diameter tube. At zero field (0 V/μm; green line), as it is approximately the case in our diodes, there is no absorption in the energy range between the $E_{11}$ exciton and the onset of the weak band-to-band absorption. The absorption, as well as the emission, are hence dominated by the $E_{11}$ excitonic transition. In the FET (25 V/μm; red line), however, due to the high electric field, the exciton wavefunction mixes with the band-to-band continuum, which leads to spectral weight transfer from the excitonic peak to the continuum. The band-to-band absorption moves into the forbidden region and merges with the $E_{11}$ exciton peak, resulting in a strongly broadened, asymmetric lineshape. At 25

V/µm, the simulated absorption extends over an energetic range of more than 150 meV. Due to the very high carrier temperatures in those devices[7], we expect the emission spectrum to be of comparable width as the absorption. Additional broadening mechanisms, such as Auger recombination[23–25] and phonon broadening[29], will increase the width even further.

Let us finally comment on the efficiency of the nanotube LEDs. Measurements of the PL efficiency of single-walled carbon nanotubes yielded values up to[25,30] $\sim 10^{-2}$, whereas in our devices we obtain at most $\sim 10^{-4}$ photons per injected electron-hole pair. This about two orders of magnitude difference can be understood by taking the following two factors into account: (i) Only a fraction of electrically induced electron-hole pairs possess the right spin to populate radiative singlet exciton states; $[1 + 3 \cdot \exp(\Delta/kT)]$ times as many populate non-radiative triplet states ($k$ is the Boltzmann constant, $T$ is the temperature, and $\Delta$ is the singlet-triplet splitting). Using literature values[13,14] for $\Delta$, we estimate that this effect reduces the efficiency by about an order of magnitude. (ii) The short non-radiative lifetime leads to an efficiency reduction by another order of magnitude. Possible routes for improving the efficiency hence would be brightening of the triplet states – for example, by adding magnetic nanoparticles[21] – and/or suspending the nanotube to increase the non-radiative lifetime.

**Methods**

The p- and n-type regions in our nanotube diodes are formed using electrostatic doping. As illustrated in Fig. 1a, two separate gate electrodes, that couple to two different regions of a single-walled carbon nanotube, are used. One gate is biased with a negative voltage, drawing holes into the nanotube channel, and the other gate is biased with a positive voltage, resulting in an accumulation of electrons in the channel. In this way, a p-n junction can be formed and the devices behave very much like conventional semiconductor diodes[11].

In a first step of device fabrication, single-walled carbon nanotubes were grown by ethanol chemical vapor deposition (3–4 nm iron oxide nanoparticles) on a highly p-doped silicon substrate with $t_{ox}$ = 200 nm thick thermal silicon oxide. The nanotubes are up to tens of micrometers long and their diameters range from 1 to 2 nm, as determined by AFM. The tube density was kept low (~1 tube per 1000 $\mu m^2$) in order to prevent the formation of nanotube bundles. Standard e-beam lithography, e-beam evaporation, and lift-off were then used to fabricate the 50-nm-thick Ti contact electrodes. Ti was chosen because it allows lineup of the Fermi level close to the middle of the nanotube band gap and hence efficient injection of both p- and n-type carriers. The devices were then annealed in vacuum and their FET characteristics were recorded by using the silicon substrate as a back-gate. The back-gated devices exhibited clearly ambipolar transfer characteristics with high on/off ratios (> $10^3$). A 33-nm-thick $Al_2O_3$ gate dielectric was then deposited on top of the sample by atomic layer deposition. Besides acting as a gate oxide, it protects the nanotube from being influenced by chemical dopants and the devices are found to be stable over months. The dielectric constant of the $Al_2O_3$ is $\varepsilon_{Al2O3}$ ~ 7.5, as determined by C-V measurements. In a second lithography step, Ti split gates were fabricated. The width of the intrinsic (ungated) region between the split gates was 1 $\mu m$. The gate field regions that produce electrostatic doping between the edges of the contact electrodes and the edges of the gates were made 1–2 $\mu m$ wide. Finally, contact windows were etched in the gate oxide at the position of the drain/source pads using phosphoric acid. A schematic drawing of the device is shown in Fig. 1a. The sample was mounted in an evacuated probe station and all measurements were performed at room temperature.

EL from the sample was collected with a 20× microscope objective and imaged onto a liquid-nitrogen cooled HgCdTe infrared camera (256 × 256 pixel; IR Laboratories). A (cooled) 2215-nm short pass filter was mounted in front of the camera to reduce the background black-body radiation and increase the sensitivity of the measurement. Emission spectra were acquired by placing a GRISM (combination of a grating and prism) in the beam path between the objective and the infrared camera. The emission lines of a xenon spectral calibration lamp were recorded to calibrate the system. The raw data were corrected for the spectral transmission of all optical components in the beam path and the spectral sensitivity of the detector.

**Acknowledgements**

We thank Zhihong Chen, Marcus Freitag, Yu-Ming Lin, Emilio E. Mendez, and Fengnian Xia for helpful discussions, and Bruce A. Ek for technical assistance. T.M. acknowledges financial support by the Austrian Science Fund FWF (Erwin Schrödinger fellowship J2705-N16).

**Author contributions**

T.M. and M.K. were responsible for the experimental work. All authors discussed the results and commented on the manuscript.

**Additional information**

Supplementary information accompanies this paper at www.nature.com/naturenanotechnology. Reprints and permission information is available online at http://npg.nature.com/reprintsandpermissions. Correspondence and requests for materials should be addressed to T.M. and P.A.

**Competing financial interests**

The authors declare that they have no competing financial interests.

**Figure captions**

Figure 1 **Device structure and electronic characteristics. a,** Schematic drawing of the carbon nanotube LED. **b,** Electrical device characteristics for different biasing conditions. Solid red line: $V_{GS1}$ = -8 V, $V_{GS2}$ = +8 V. The nanotube is operated as a diode and shows rectifying behavior. Dashed green line: $V_{GS1}$ = $V_{GS2}$ = -8 V. The nanotube behaves as p-type resistor. The silicon bottom-gate was grounded during the measurements. **c,** Bandstructure of the nanotube diode when it is biased in forward direction ($V_{DS} > 0$). Electrons and holes are injected into the intrinsic region and recombine partially radiatively and partially non-radiatively.

Figure 2 **Identification of the light emission mechanism. a,** The upper plane is an SEM image of a carbon nanotube p-n diode. The nanotube is shown in black, the drain (D) and source (S) electrodes are yellow, and the overlapping gate electrodes are in green/yellow. Note that the rightmost electrode and gate belong to a different device and are not contacted. The scale bar is 10 µm long. The lower plane is a surface plot of the infrared emission. A microscopy image of the device (not shown) was taken under external illumination in order to verify that the emission is localized at the position of the tube. Infrared emission is observed at the position of the tube when the device is operated as a LED ($V_{GS1}$ = -8 V, $V_{GS2}$ = +8 V, $I_{DS}$ = 240 nA; left image). In contrast, no emission is observed when a unipolar current of equal magnitude is driven through the nanotube ($V_{GS1}$ = -8 V, $V_{GS2}$ = -8 V, $I_{DS}$ = 240 nA; right image). **b,** EL spectrum of a nanotube diode at $I_{DS}$ = 200 nA.

Figure 3 **Electroluminescence spectra. a,** EL spectrum of a nanotube diode recorded at different drain-source currents $I_{DS}$ between 30 and 230 nA. Gate biases of $V_{GS1}$ = -9 V and $V_{GS2}$ = +9 V were applied. The data can be fitted well with two Gaussians and, at low currents, we extract widths of ~45 meV (FWHM) for the individual contributions. Besides the strong exciton emission (labeled *X*), a weaker satellite peak at lower energy is observed (*LX*). It is attributed to localized exciton emission. **b,** Comparison between EL spectra at two different gate biases (normalized). Solid green line: $V_{GS1}$ = -7 V, $V_{GS2}$ = +7 V; dashed grey line: $V_{GS1}$ = -9 V, $V_{GS2}$ = +9 V. The spectral width of the -/+ 7 V measurement is only ~35 meV (FWHM). **c,** Red symbols: The free exciton emission (*X*) shows an approximately linear increase with current. Green symbols: Localized exciton emission (*LX*). The EL saturates as the current exceeds ~100 nA. The EL versus current dependence shows no threshold behavior. The dashed lines are guides to the eye. **d,** Same as **c**, but for a different device.

Figure 4 **Comparison between ambipolar and unipolar light emission.** Green line: EL of a nanotube diode. Red line: EL emission from a FET device. The upper curve is offset for clarity. Both devices are made out of the same, long tube and undergo the same processing steps with the only difference being the absence of top-gate electrodes in the FET device. Inset: Calculated bandstructure[27] of the FET. In the FET emission measurements, biases of $V_{DS}$ = +5 V and $V_{GS}$ = -20 V were applied to the drain- and (silicon) back-gate electrodes, respectively. An electric field of >25 V/μm occurs near the drain electrode. Holes are injected from the drain (i), accumulate kinetic energy in the high electric field (ii), and eventually generate electron-hole pairs by impact excitation (iii). Green line: Calculated absorption spectrum of a 1.4 nm diameter tube ($\varepsilon_{eff}=6.0$) at zero field. *X* is the exciton transition, *FC* denotes the band-to-band (free carrier) transitions. Red line: Calculated absorption spectrum for a field of 25 V/μm. For direct comparison, we also overlay the calculated absorption with the measurement in the main panel. Inhomogeneous broadening was taken into account by convoluting the calculated spectra with Gaussians of 25 meV FWHM (inset) and 50 meV FWHM (main panel) widths. The simulation does not reproduce the experimentally observed blue-shift of the FET emission with respect to *X*. This shift most likely arises from different dielectric environments in the two devices.

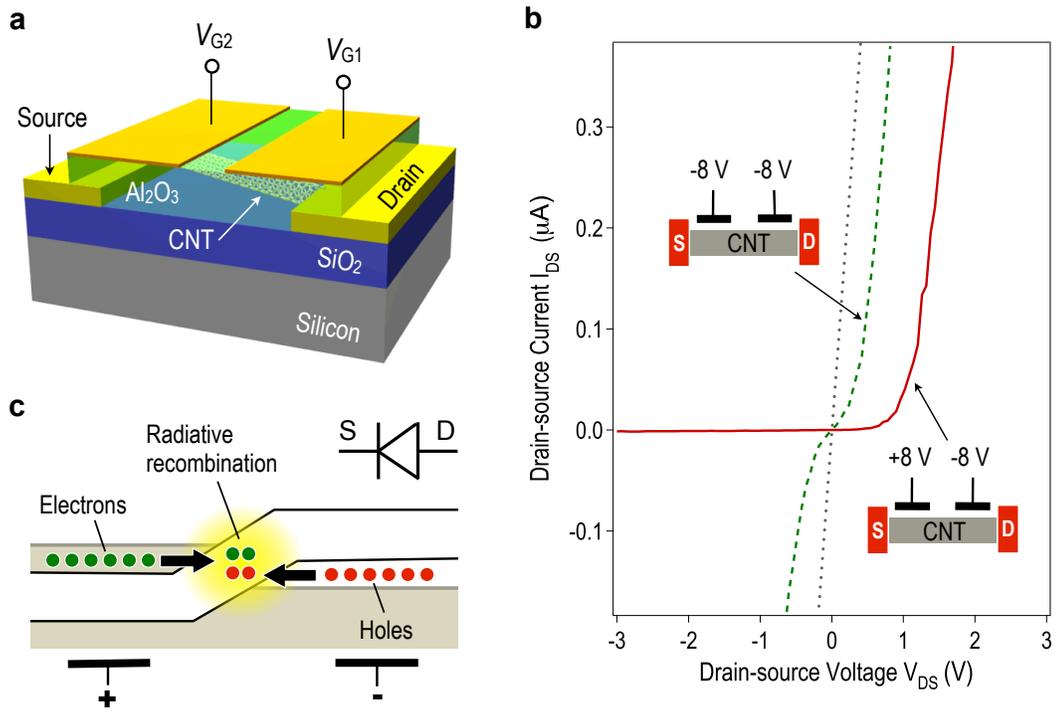

Figure 1

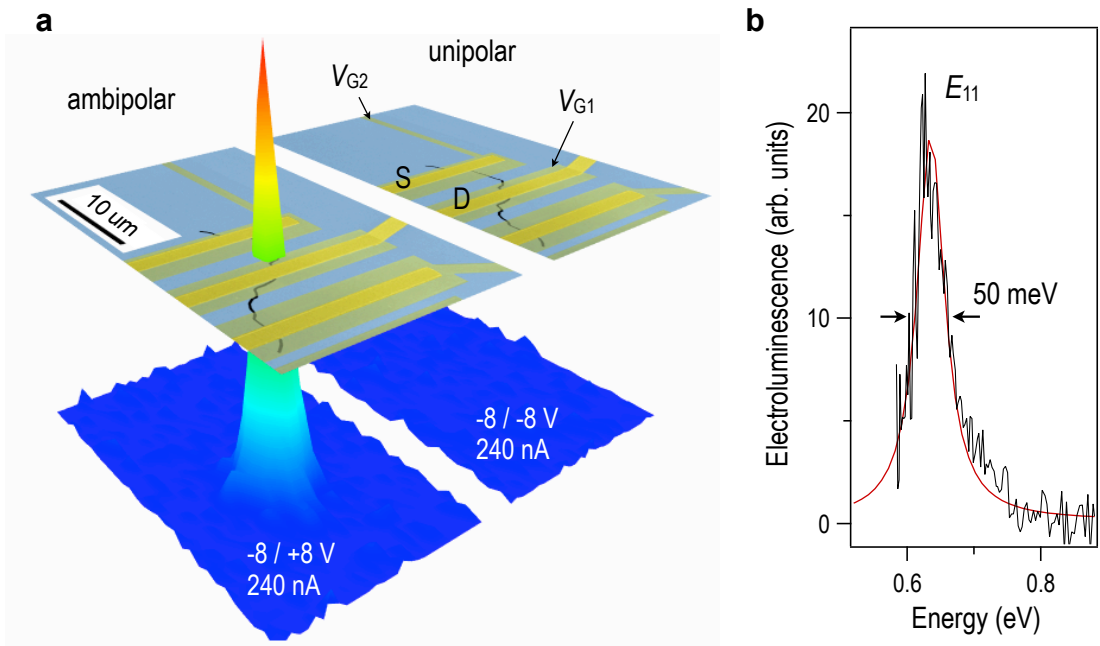

Figure 2

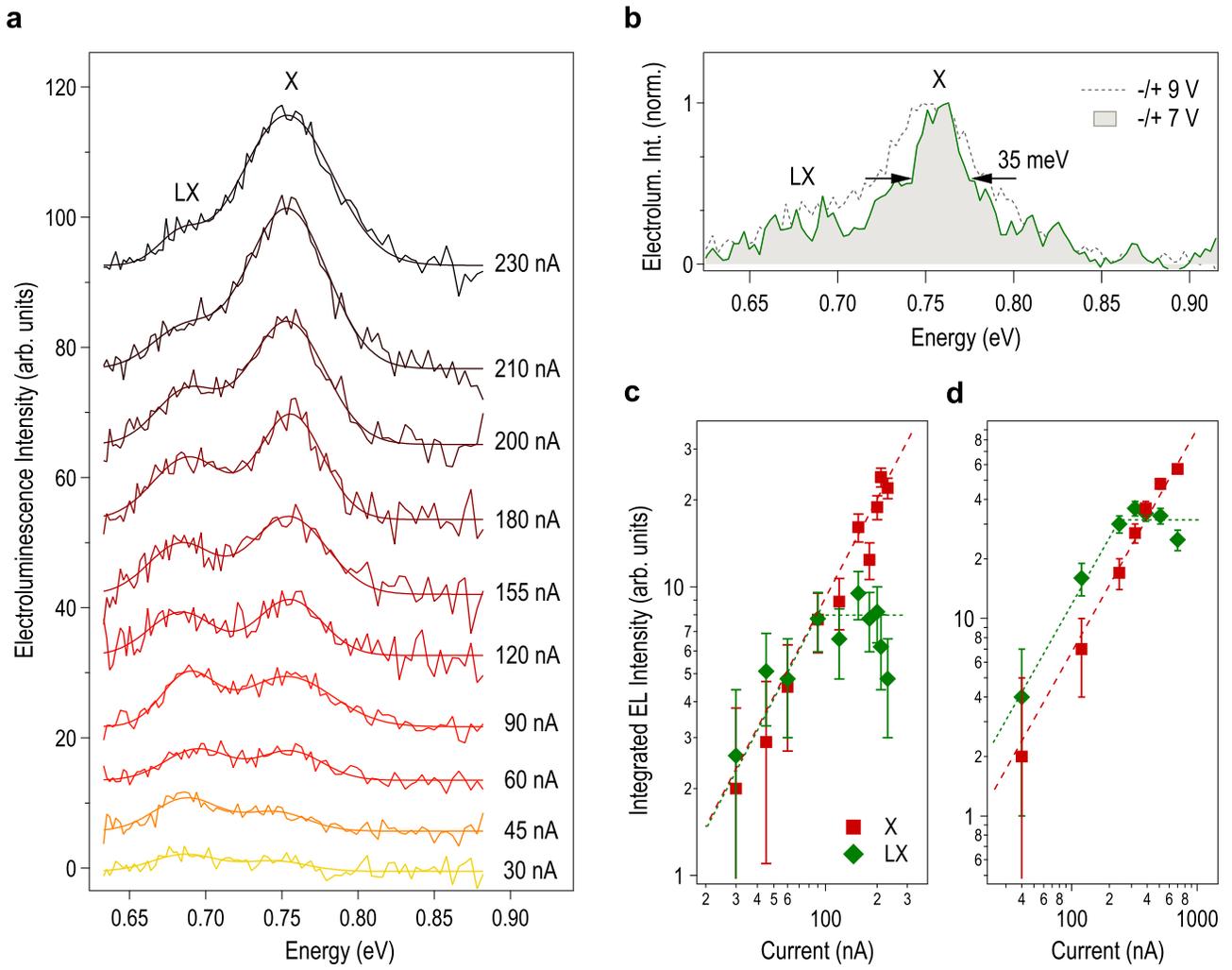

Figure 3

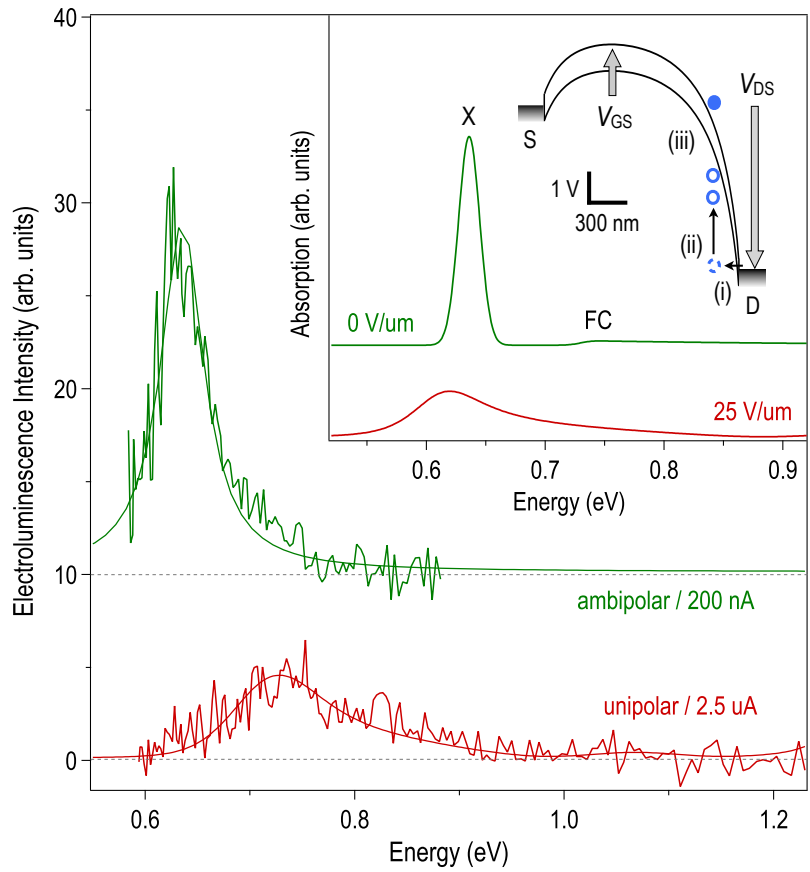

Figure 4